\documentclass[10pt,conference]{IEEEtran}

\usepackage[utf8]{inputenc} 
\usepackage[T1]{fontenc}    
\usepackage{hyperref}       
\usepackage{url}            
\usepackage{booktabs}       
\usepackage{nicefrac}       
\usepackage{microtype}      
\usepackage[table]{xcolor}  
\usepackage{comment}
\usepackage{xspace}
\usepackage{listings,multicol}
\usepackage[caption=false]{subfig}  
\usepackage[export]{adjustbox}
\usepackage{amsmath}
\usepackage{amssymb}
\usepackage{textcomp}
\usepackage{breakurl}
\usepackage{algorithm}
\usepackage[noend]{algpseudocode}
\usepackage{cleveref}
\usepackage{multirow}
\usepackage{enumerate}
\usepackage[shortlabels]{enumitem}
\usepackage{makecell}
\usepackage{spverbatim}
\usepackage{fancyvrb}
\usepackage{tabularx}
\usepackage{tablefootnote}
\usepackage{graphicx}
\usepackage{soul}
\usepackage{balance}
\usepackage{comment}

\newcommand{\benchmark}{PandasPlotBench}

\title{Drawing Pandas: A Benchmark for LLMs \\ in Generating Plotting Code}

\author{\IEEEauthorblockN{Timur Galimzyanov\IEEEauthorrefmark{1}, Sergey Titov\IEEEauthorrefmark{1}, Yaroslav Golubev\IEEEauthorrefmark{1}, Egor Bogomolov\IEEEauthorrefmark{2}\IEEEauthorrefmark{1}}
\IEEEauthorblockA{\IEEEauthorrefmark{1}\textit{JetBrains Research}, \IEEEauthorrefmark{2}\textit{Delft University of Technology} \\
\{timur.galimzyanov, sergey.titov, yaroslav.golubev, egor.bogomolov\}@jetbrains.com
}
}

\begin{document}

\maketitle

\begin{abstract}

This paper introduces the human-curated \emph{\benchmark} dataset, designed to evaluate language models' effectiveness as assistants in visual data exploration. Our benchmark focuses on generating code for visualizing tabular data---such as a Pandas DataFrame---based on natural language instructions, complementing current evaluation tools and expanding their scope. The dataset includes 175 unique tasks. Our experiments assess several leading Large Language Models (LLMs) across three visualization libraries: Matplotlib, Seaborn, and Plotly. We show that the shortening of tasks has a minimal effect on plotting capabilities, allowing for the user interface that accommodates concise user input without sacrificing functionality or accuracy. Another of our findings reveals that while LLMs perform well with popular libraries like Matplotlib and Seaborn, challenges persist with Plotly, highlighting areas for improvement. We hope that the modular design of our benchmark will broaden the current studies on generating visualizations. Our dataset and benchmark code are available online: \url{https://huggingface.co/datasets/JetBrains-Research/PandasPlotBench}; \url{https://github.com/JetBrains-Research/PandasPlotBench}.

\end{abstract}

\section{Introduction}\label{sec:introduction}

Visualizing data is crucial in data analysis, with visual representations being essential for uncovering patterns and trends that are not immediately apparent from raw data alone.

Recent research demonstrates the growing use of Large Language Models (LLMs) for code generation in data analysis and visualization tasks. LLMs can automate various stages of the data analysis pipeline, from processing data to creating visualizations and interpreting results~\cite{Jansen2023LeveragingLLMs, Beasley2024Pipe, Nejjar2023LLMs, matplotagent}. While LLMs show promise in generating code for data visualization and analysis, challenges remain in producing fully executable code for complex tasks~\cite{Jansen2023LeveragingLLMs, Vaithilingam2022Expectation}. These challenges are compounded by a lack of robust benchmarks to evaluate a model's ability to write accurate plotting code based on brief instructions and data descriptions.

The domain of data visualization benchmarks for LLMs encompasses various tasks, including generating concise Matplotlib code snippets~\cite{ds-1000}, visually guided code generation~\cite{matplotagent, plot2code, chartmimic}, as well as data analysis tasks leveraging SQL queries and visualization languages~\cite{nvbench}. However, only one work, MatPlotBench~\cite{matplotagent}, addresses the task of visualizing data from a given DataFrame or a CSV file, although it comprises only a limited number of data points (25 datapoints using CSV data files). Moreover, current benchmarks fall short in providing an in-depth examination of how rephrasing instructions and utilizing different plotting libraries can impact performance, as well as benchmarking across a wide range of models.

In this work, to overcome this existing gap, we introduce the human-curated \emph{\benchmark}, designed to assess AI models and approaches as assistants in visual data exploration. Our primary focus is on generating code for plotting data loaded in the \textit{Pandas DataFrame} \cite{pandas} format based on task instructions, for which we provide a robust and comprehensive dataset comprising 175 unique and well-defined tasks. Our benchmark aims to be free from data leakage, as the tasks and data files are synthetically generated.

Our benchmark aims to provide valuable insights that can help researchers improve user experience in data visualization and analysis. To investigate various real-world scenarios and showcase the benchmark, we conducted a series of experiments to evaluate the performance of LLMs in generating visualizations based on user instructions. We tested several popular LLMs, including OpenAI GPT-4o~\cite{gpt4o}, Anthropic Claude 3.5 Sonnet, 3 Opus, 3 Haiku~\cite{claude3}, Google Gemini 1.5 Pro~\cite{gemini}, and the Llama family of models~\cite{dubey2024llama}, examining how well each model could interpret DataFrame descriptions and translate them into accurate visualizations.

One key area of focus was the effect of task compression (\textit{i.e.}, summarization of instructions) on model performance. We hypothesized that summarizing tasks into a single sentence could help mimic user behaviour, as real-world users often provide concise instructions. Our results indicated that task compression had a minimal impact on the models' plotting capabilities when the models were provided with well-structured DataFrame descriptions.

Our results also highlight the difference in models' proficiency across various visualization libraries. Specifically, we benchmarked Matplotlib \cite{matplotlib}, Seaborn \cite{seaborn}, and Plotly \cite{plotly}, and predictably found that models struggle to generate code with the underrepresented Plotly library.

The benchmark is available online~\cite{hf}, and we also open-source our implementation of the baseline methods and the evaluation code, available on our GitHub page~\cite{github}. For the full list of instruction prompts used for processing the data, generating the results, and scoring, as well as the demonstration of all the tasks, scores and generated plots, please refer to the supplementary materials~\cite{supplementary}.

\section{\benchmark}
\label{sec:dataset}

\subsection{Data Collection and Processing}

To construct the benchmark, we sourced data from the Matplotlib gallery~\cite{gallery}, which contains 501 scripts for creating various plots for the hard-coded data. We converted these scripts into data points, each comprising a Pandas DataFrame and a task prompt for plotting. The process was as follows:

\begin{enumerate}
    \item \textbf{Filtering valid files}. Executed the gallery scripts in a Jupyter environment to select only those files that produced valid plots. This reduced the data points to 307.
    \item \textbf{Splitting plotting code}. Using OpenAI GPT-4 \cite{gpt4}, we split each script into two components: (1) a data-generating part that generates the plotting data and consolidates  into a single Pandas DataFrame, and (2) a plotting part that takes the DataFrame as input.
    \item \textbf{Manual verification}. We manually reviewed each split to ensure accuracy, correcting errors where possible, and saved the resulting DataFrames as CSV files. After this step, 201 data points remained.
    \item \textbf{Final validation}. We manually verified that each code segment correctly plots the data from its associated CSV file, resulting in 175 final data points.
    \item \textbf{Generating plotting tasks}. We used GPT-4V~\cite{gpt4v} to create detailed instructions for plotting each DataFrame, based on the plotting code, the DataFrame metadata, and the plot image. This process included the manual verification of all the tasks, as well as generating shorter versions of tasks: a short one (2-3 sentences) and a single-sentence one, to later test their performance.
\end{enumerate}

The prompts for splitting the code and generating the tasks can be found in the supplementary materials~\cite{supplementary}. The dataset is available on the HuggingFace page~\cite{hf}, the benchmark code is available on GitHub~\cite{github}.

\subsection{Data Characteristics}

The final dataset consists of 175 data points designed to simulate typical user tasks in data visualization. We assume that a user possesses data in the format of a Pandas DataFrame and wishes to create a plot by describing the task to the model. Each data point includes the following elements:

\begin{enumerate}

    \item A CSV file containing the data to be plotted.
    \item A script for loading and preparing the data, which may include minimal data processing steps (\textit{e.g.}, converting string fields to objects). For most data points, this script is: \textit{import pandas as pd // df = pd.read\_csv("data.csv")}

    \item A ground truth plot.
    \item Ground truth code that visualizes the pre-loaded Pandas DataFrame using Matplotlib.
    \item Three versions of a plotting task to guide the model:
        \begin{itemize}
            \item A detailed task prompt for plotting.
            \item A 2-3 sentences compressed version of the task.
            \item A single-sentence compressed version of the task.
        \end{itemize}
\end{enumerate}

The detailed task is split into two sections: \textbf{Plot description}, which provides details of the target plot, and \textbf{Plot style} description, which offers general guidelines for plot styling. When benchmarking on compressed tasks, the plot style is omitted. For an example of a task from a random data point, a system prompt, and plotting instructions, please refer to the supplementary materials~\cite{supplementary}.

\subsection{Evaluation Metrics}

Using the provided model, our benchmark generates the plotting code according to the given task. This code is then executed within a Jupyter Notebook environment, which is subsequently parsed to obtain the plot image. We chose Jupyter as a universal environment that enables seamless switching between plotting libraries and easy parsing of plot images. Each plotting code is executed in an individual notebook cell. We track the ratio of \textit{incorrect code}, defined as cells that did not result in a plot. When the notebook cells return both plots and errors, we ignore the errors. The scoring is performed by the multimodal GPT-4o Judge model in two ways:

\textbf{Visual scoring}. The Judge model is asked to compare the generated and the ground truth plots on 0--100 scale, focusing on the main idea of the plots.

\textbf{Task-based scoring}. The Judge model is asked to score the adherence of the resulting plot to the task description.

It is important to note that the generated images do not have to match the ground-truth plots exactly, as the tasks were provided as text and are not extremely detailed. Additionally, the dataset tasks were generated by the GPT-4V model based on the given code and plot image. Therefore, some information may have been lost during this two-step process. In our opinion, the generated plots were often better than the ground-truth ones, leading us to include and prefer task-based scoring, which generally yields higher scores on average. Moreover, the Pearson correlation between task-based scores and visual scores for the best-performing model is 0.58, highlighting the importance of using both approaches. Judging prompts are provided in the supplementary materials~\cite{supplementary} and are based on the corresponding prompts from the work of Yang et al.~\cite{matplotagent}.

We also evaluated the equivalence of ground truth and generated plotting code using the CodeBERT Score~\cite{codebert}, but found it inadequate for assessing code quality and decided not to use it. It showed no correlation with visual or task scores, as both correct and incorrect code targeted the same task and differed only in implementation details. 

To ensure the robustness of our benchmark, we performed the preliminary human scoring of the results of the best-performing generating setup and analyzed its correlation to the two scores above. The first author of the paper, with 5 years of experience in Python and data visualization, went through all the data points and manually scored them judging the adherence of the resulting plot to the task description. The analysis revealed a strong correlation between the human scores and task-based scores, with a Pearson correlation coefficient of 0.85, suggesting close alignment. In contrast, the correlation between human scores and visual scores is notably lower, with a Pearson coefficient of 0.66, indicating a moderate relationship. This quick check allows us to trust in the meaningful results of the benchmark, however, it is crucial to carry out a thorough investigation of the scores correlating with human judgement in future work, with multiple experts.

\newcolumntype{C}[1]{>{\centering\arraybackslash}m{#1}} 
\section{Experiments}
\label{sec:experiments}

To showcase the usefulness of the benchmark and the different important aspects that it can help study, we ran it in various production-important setups. First, we used a benchmark to determine the best way to include the DataFrame description in the prompt. Then, we benchmarked proprietary and open models to assess their capabilities in generating plotting code (see Section~\ref{sec:exp-models}). The best-performing model (GPT-4o) was used for subsequent experiments. 

We then benchmarked the three most popular plotting libraries: Matplotlib, Seaborn, and Plotly (see Section~\ref{sec:exp-plot-lib}). Given that users are generally not inclined to write thoroughly detailed plotting instructions, we explored the effect of task length on the quality of the generated plots (see Section~\ref{sec:exp-task_len}). For all experiments, we ran the benchmark five times for each factor to ensure the robustness of our results.

To increase the speed of code generation, we explicitly instructed the model to write fewer explanations and return only the code within a code block with comments (see plotting instructions in supplementary materials~\cite{supplementary}). This modification did not significantly affect the benchmark scores.

For all experiments, we highlight the following results for each model: the mean score on a 0--100 scale (one for \textbf{visual} score and one for \textbf{task-based} score) and the ratio of \textit{good} scores among all ones (\textit{i.e.}, how many of the data points received a score $\geq$75, again separate for the two scores).

\subsection{Benchmarking Models}
\label{sec:exp-models}

First, to set up model prompt construction, we experimented with different DataFrame descriptions and found that the \textit{head(5)} method (\textit{i.e.}, the first five lines), supplemented with column names and types, delivered the best performance. For examples, refer to the supplementary materials~\cite{supplementary}.

We benchmarked the following proprietary models: OpenAI GPT-4o~\cite{gpt4o}, Anthropic Claude 3.5 Sonnet, 3 Opus, 3 Haiku~\cite{claude3},  and Google Gemini 1.5 Pro~\cite{gemini}, as well as Llama instruct models~\cite{dubey2024llama}: Llama 3.2 (1B, 3B) and INT8 Llama 3.1 (8B, 70B, 405B). The results are summarized in Table~\ref{tab:scores-models}.
GPT-4o and Claude 3.5 Sonnet models share the highest scores. Sonnet performs 20\% slower, but this is due to longer responses, which could be mitigated through prompt tuning to enforce shorter answers. The fastest proprietary model is the Claude 3 Haiku, however, it has a significantly lower score. Importantly, Anthropic Claude models tend to generate excessively long responses, even when explicitly instructed to be concise. For further experiments, we chose to use GPT-4o, as it is the fastest among the highest-performing models. 

Large Llama models (70B and 405B) exhibit performance comparable to proprietary LLMs, while smaller Llama models often fail to produce functional code. Perhaps, optimizing the prompts might improve their performance.

\begin{table}[t]
\resizebox{\columnwidth}{!}{%
\begin{tabular}{@{}lcccccc@{}}
\toprule
\multirow{2}{*}{Model} &
  \multirow{2}{*}{\begin{tabular}[c]{@{}l@{}}Incorrect \\ code, \%\end{tabular}} &
  \multirow{2}{*}{\begin{tabular}[c]{@{}l@{}}Time \\ (s/item)\end{tabular}} &
  \multicolumn{2}{c}{Mean score} &
  \multicolumn{2}{c}{Good ($\geq$75)} \\ \cmidrule(lr){4-5} \cmidrule(lr){6-7} 
                  &      &      & visual & task & visual        & task \\ \midrule
GPT-4o            & 1.8  & 4.8  & 75     & 89   & \textbf{0.69} & \textbf{0.91} \\
Claude 3.5 Sonnet & 2.5  & 5.8  & 73     & 88   & \textbf{0.69} & 0.89 \\
Claude 3 Opus     & 3.0  & 14.0 & 72     & 87   & 0.65          & 0.89 \\
Gemini 1.5 Pro    & 1.7  & 9.8  & 71     & 81   & 0.64          & 0.81 \\
Claude 3 Haiku    & 2.9  & 3.7  & 65     & 78   & 0.60          & 0.77 \\ \midrule
Llama 3.1 405B    & 2.9  & 5.7  & 73     & 86   & 0.66          & 0.88 \\
Llama 3.1 70B     & 4.6  & 4.2  & 68     & 82   & 0.62          & 0.82 \\
Llama 3.1 8B      & 7.4  & 3.0  & 56     & 68   & 0.49          & 0.66 \\
Llama 3.2 3B      & 4.6  & 1.9  & 45     & 62   & 0.36          & 0.56 \\
Llama 3.2 1B      & 14.9 & 0.3  & 34     & 40   & 0.22          & 0.30 \\ \bottomrule
\end{tabular}%
}
\caption{Benchmark scores for different models.}
\label{tab:scores-models}
\vspace{-0.4cm}
\end{table}

\subsection{Benchmarking Plotting Libraries}
\label{sec:exp-plot-lib}
To evaluate the model's ability to generate code across different plotting libraries, we requested GPT-4o to produce code for Matplotlib, Seaborn, and Plotly (see Table~\ref{tab:scores-libs}). We explicitly instructed the model to use the target library and not others. For the generated code, we checked whether the target library was imported. If the target library was not imported, the score was set to 0, and such errors were aggregated in the \textit{"Wrong library"} column.
Both Matplotlib and Seaborn libraries demonstrated high scores. Since the ground truth for visual scoring was always the Matplotlib score, the visual-guided scoring was predictably lower for other libraries.
Still, Plotly was much less familiar to GPT-4o, resulting in 38 instances (22\%) of incorrect code. We manually reviewed these cases and found that most errors were due to the incorrect usage of the Plotly library API by the language model. 

\begin{table}[t]
\resizebox{\columnwidth}{!}{%
\begin{tabular}{@{}lcccccc@{}}
\toprule
\multirow{2}{*}{Library} &
  \multirow{2}{*}{\begin{tabular}[c]{@{}l@{}}Wrong \\ library\end{tabular}} &
  \multirow{2}{*}{\begin{tabular}[c]{@{}l@{}}Incorrect\\ code, \%\end{tabular}} &
  \multicolumn{2}{c}{Mean score} &
  \multicolumn{2}{c}{Good ($\geq$75)} \\ \cmidrule(lr){4-5} \cmidrule(lr){6-7} 
  &      &      & visual & task & visual        & task \\ \midrule
Matplotlib & 0 & 1.8  & 75&89 & 0.69&0.91 \\
Seaborn    & 1 & 5.2  & 67&84 & 0.60&0.86 \\
Plotly     & 0 & 22.0 & 59&68 & 0.55&0.70 \\ \bottomrule
\end{tabular}%
}
\caption{Benchmark scores for different plotting libraries.}
\label{tab:scores-libs}
\vspace{-0.7cm}
\end{table}

\subsection{Alternating Task Length}
\label{sec:exp-task_len}

To explore the effect of task length on the quality of the plots, we tested the best-performing GPT-4o model with different versions of tasks included in our dataset. As a control, we replaced the task with a generic instruction: "\textit{Plot a given dataframe}." In Table~\ref{tab:task-len-scores}, it can be seen that the shortened versions are 2-3 times shorter than the basic task. Note that this reduction affects only the task part of the prompt, while other instructions and data descriptions remain intact, averaging around 1600 symbols. Therefore, the primary motivation for trying shorter tasks is to test potential improvements in user experience rather than significant cost reduction. For task-based scoring, we used the original task in the judging request to maintain consistency with previous results.

The results (see Table~\ref{tab:task-len-scores}) demonstrate that even a significant shortening of the task description only slightly decreased the scores. In contrast, en the task was omitted and replaced with a generic instruction, only one-third of the plots could be considered satisfactory. Therefore, we conclude that even a single-sentence task description written by a user can produce positive results, provided that automatically generated DataFrame descriptions and thorough instructions are given.

\begin{table}[t]
\resizebox{\columnwidth}{!}{%
\begin{tabular}{@{}lccccccc@{}}
\toprule
\multirow{2}{*}{Plot task} &
  \multirow{2}{*}{\begin{tabular}[c]{@{}l@{}}Task length\\ (symbols)\end{tabular}} &
  \multirow{2}{*}{\begin{tabular}[c]{@{}l@{}}Incorrect\\ code, \%\end{tabular}} &
  \multirow{2}{*}{\begin{tabular}[c]{@{}l@{}}Time\\ (s/item)\end{tabular}} &
  \multicolumn{2}{c}{Mean score} &
  \multicolumn{2}{c}{Good ($\geq$75)} \\ \cmidrule(lr){5-6} \cmidrule(lr){7-8} 
  &      &      & & visual & task & visual        & task \\ \midrule
Basic           & 736 & 1.8 & 4.9 & 75&89 & 0.69&0.91 \\
No style        & 355 & 1.1 & 4.5 & 74&86 & 0.69&0.85 \\
Short           & 201 & 1.4 & 4.2 & 72&84 & 0.66&0.83 \\
Single sentence & 154 & 0.2 & 4.3 & 71&85 & 0.64&0.85 \\
No task         & 22  & 0.6 & 4.2 & 44&36 & 0.32&0.23 \\ \bottomrule
\end{tabular}%
}
\caption{Benchmark scores for different task formulations.}
\vspace{-0.5cm}
\label{tab:task-len-scores}

\end{table}
\section{Related Work}
\label{sec:relaetd-work}

Our work closely follows the methodologies established by MatPlotBench \cite{matplotagent} and Plot2Code \cite{plot2code}. Our data point structure inherits key elements from MatPlotBench, specifically the inclusion of data, task, and ground-truth plot components. However, despite MatPlotBench's claim of offering 100 data visualization problems “closely aligned with real-world scenarios”, only 25 include actual data files, while the rest rely on natural-language plotting instructions. Plot2Code, in turn, provides 132 plots with corresponding tasks, but it relies on natural-language instructions and lacks data files. Furthermore, it primarily focuses on image-guided plotting code generation, which does not reflect the primary real-world use case.

A closely related dataset is ChartMimic \cite{chartmimic}, which includes 1,000 human-curated examples focused on generating code based on provided plot images. This dataset offers an option to include Python-embedded data in the prompt, such as defining lists of data and column names directly in the code. Another analogue is nvBench \cite{nvbench}, which comprises 153 data files and includes 7,274 visualizations. Each visualization in nvBench is paired with one or more natural language queries. However, nvBench targets more complex tasks, including data analysis, and focuses on SQL queries and Vega-Lite-like languages for visualization. It is also important to mention the DS-1000 benchmark \cite{ds-1000}, which consists of a thousand data science problems, including coding challenges involving Matplotlib. However, the solutions in DS-1000 average merely three lines in length, making them too simplistic for assessing modern LLMs and AI agents in solving practical, real-world problems.

While the datasets and benchmarks mentioned above fulfil the role of evaluating the plot-building capabilities for the most popular Python frameworks, they lack extensibility. To address this limitation, our evaluation framework enables users to assess the plot-building capabilities of models across different styles, frameworks, and even programming languages. To achieve this, in addition to the base dataset of tasks and data that we present, our methodology can be used for generating code in the selected framework or language. Furthermore, we structure our tasks to distinguish between parts of the prompt that (1) define setup and framework, (2) task, and (3) styling options. We believe that this approach will extend LLM evaluation beyond just Matplotlib and Python.
Moreover, our approach not only aids in evaluating plotting capabilities but also facilitates experimentation with LLM-based plot building. To highlight this, in this paper we presented several experiments exploring the effects of a library choice and task verbosity on plot creation.

\section{Limitations and Future Work}
\label{sec:limitations}

The benchmark is based on open-source and synthetically generated plotting data. Although the synthetic tasks and data files have never been seen by any LLMs, models may still exhibit bias towards publicly available ground-truth Matplotlib code. However, this code has been modified, and solutions using other libraries remain unseen. Another potential issue could be a bias towards OpenAI models, which were used for synthetic task generation, but we believe this bias to be negligible. Therefore, we argue that our tasks are less prone to models memorizing training data, as there is no direct link between the answers to benchmark tasks and the raw repository data that modern models use for training.

Our dataset is currently limited to 175 data points originating from the Matplotlib gallery. Future work should include expansion to other sources. Due to this limitation, we used the Matplotlib ground truth plots to evaluate other libraries, relying more on the \textit{task-based scoring}. Also, the DataFrames in the dataset are mostly concise, containing only the data required for plotting, often far from real-world data. Enriching, augmenting, and modifying these files with random columns and additional data could make the generation task more representative of real-world scenarios and more complex.

To allow for manual examination of the collected data and to maintain consistency in the benchmarks, we limited ourselves to Python code, which has become almost a standard in data analysis. We plan to extend the datasets to other languages in future work. Also, the manual scoring of the data points was carried out only by the first author, as a general sanity check of the LLM judging quality. In the future, we plan to conduct more thorough expert validation.  

\section{Conclusion}

In this paper, we present the human-curated \emph{\benchmark} dataset to provide an assessment of AI models as assistants in visual data exploration. Our benchmark focuses on plotting the data from a Pandas DataFrame and is free from data leakage by virtue of being synthetically generated.

To showcase the diversity of our benchmark, we tested numerous LLMs in various real-world scenarios: with different ways of describing a DataFrame, different lengths of tasks, and different plotting libraries. While for the most popular Matplotlib and Seaborn libraries, LLMs are capable of producing compilable and high-scoring code, their knowledge of Plotly is still far from perfect with approximately 22\% of failed code cases. As for the task length, we show that given automatically generated descriptions of the DataFrame, significant task shortening does not substantially affect plotting capability, which is promising for further UI development. 

Although state-of-the-art proprietary LLMs and large open models (Llama 3.1 70B and 405B) demonstrate high scores, there is still room for improvement, especially in working with less popular visualization libraries. We hope that our benchmark will assist researchers in enhancing user experience in data visualization and analysis.

\bibliographystyle{IEEEtran}
\balance
\bibliography{paper}

\newpage

\end{document}